\documentclass[11pt]{article}
\usepackage{amssymb,amsmath,amsfonts}
\usepackage{graphicx}
\usepackage{graphics}
\usepackage{epsfig}

\begin{document}

\title{Exact Green's function for fermions in an external Yang-Mills gauge
field}
\author{V. V. Parazian\thanks{%
E-mail: vparazian@gmail.com } \\
%EndAName
\textit{Institute of Applied Problems of Physics NAS RA,}\\
\textit{25 Nersessian Street, 0014 Yerevan, Armenia}}
\maketitle

\begin{abstract}
We obtain the Green's function for fermions in an external non-Abelian gauge
field with an $SU(N)$ symmetry group. As an external field, we examine the
solution to the Yang-Mills equation presented as a plane wave on the light
cone.
\end{abstract}

\section{Introduction}

The study of the properties of fermion fields in external classical fields
is a significant area of research. In the papers \cite%
{Schwinger1951,Volkov1935,Ritus1985}, it is argued that exact backgrounds
are manageable and physically meaningful. They used the proper-time
technique, exact Dirac wave functions in a plane wave (the basis of all
plane-wave propagators), and the eigenfunction method in plane waves to
construct exact propagators and loop effects in intense fields. The clear
proofs used to construct exact Green's functions in plane waves are given in
\cite{Piazza2018}. The exact propagator's quasiclassical form, which is
useful for approximations and asymptotics, is studied in \cite{Piazza2022}.
A complete review of external-field methods, loops, and resummations is
provided in \cite{Fedotov2023}.

The authors \cite{Adamo2019} precisely defined non-Abelian plane waves and
derived background Feynman rules, which are essential for exact-background
expansions in YM. The first explicit calculation of a non-Abelian plane-wave
amplitude using exact background Feynman rules was performed in \cite%
{AdamoIlderton2019}. The connection between gauge theory and gravity
amplitudes was shown in \cite{AdamoIlderton2020} using the double copy
method on plane wave backgrounds.

The background field method is employed to include multi-loop calculations
in gauge field theories, as discussed in \cite{Abbott1981}. A relationship
between the effective action in the standard functional method and the
gauge-invariant effective action derived with this approach is established.
The covariant diagrammatic technique for the effective action in gauge
theories is outlined in \cite{Barvinsky1985}. A generalized version of the
Schwinger-DeWitt method is developed that can be used to calculate any
covariant loop diagram in the effective action. Useful information on heat
kernel coefficients found in mathematical and physical literature is
compiled in \cite{Vassilevich2003}. On manifolds with and without
boundaries, under local and nonlocal boundary conditions, and in the
presence of different singularities (e.g., domain walls), the author gives
explicit formulas for these coefficients. The link between standard
perturbation theory methods and the Bern-Kosower rules for gauge theory is
explained in \cite{Strassler1992}. Bern-Kosower-type rules are proposed for
one-loop effective actions with vector bosons, Dirac spinors, and scalars in
a background gauge field without utilizing Feynman diagrams or string
theory. The "string-inspired" concept of the scope and applications of
perturbative quantum field theory is summarized in \cite{Schubert2001}. The
suggested formalism skips many of the procedures of ordinary
second-quantized field theory and permits the computing of effective actions
and S-matrix elements, similar to string perturbation theory. As shown in
\cite{Dunne2021}, the one-loop Euler-Heisenberg QED effective Lagrangian in
a constant background field results in a significantly different
nonperturbative trans-series structure at two-loop and higher-loop orders in
the fine structure constant. In the article \cite{Brown1979}, the authors
were among the first to discuss quark propagation in a classical color
field. Dittrich and Reuter \cite{Dittrich1985} extended the proper-time
method to color backgrounds, providing one-loop effective actions. Ilderton
et al. (see, for instance, \cite{Ilderton,Ilderton2014}) developed methods
for computing nonperturbative propagators in non-Abelian plane-wave
backgrounds using path integrals. The authors \cite{Cucchieri2006} have
enhanced the understanding of Green's functions in Landau-gauge Yang-Mills
theory and their implications for nonperturbative phenomena in QCD. The
authors \cite{Sadri}, working in the area of plane-wave/Super Yang-Mills
duality, are actively researching gauge-invariant field correlators and the
properties of Green's functions in unconventional backgrounds, which support
both theoretical and phenomenological studies. The author \cite{Koshelkin}
has presented a quasi-classical gauge theory model and precise solutions for
Dirac fermions in non-Abelian contexts.

While Green's function for fermions in free space or external Abelian
(electromagnetic) fields is analytically easily handled, the exact treatment
in non-Abelian gauge field backgrounds remains highly nontrivial due to the
self-interacting nature of the gauge field. In this work, we address this
defiance by deriving the exact fermionic Green's function in the presence of
an external classical Yang-Mills gauge field with symmetry group $SU(N)$.

Choosing the gauge field in the form $A_{+}^{a}\left( x\right) =f^{a}\left(
x^{+}\right) x^{1}+g^{a}\left( x^{+}\right) x^{2}+h^{a}\left( x^{+}\right) $%
, with $x^{+}=x^{0}+x^{3}$ being the light-cone coordinate (see below),
makes the problem suitable for exact analytic analysis. The form of the
gauge field as a plane wave on the light cone enables the construction of
the fermion propagator, yielding a compact expression that captures the
interaction with the non-Abelian background.

Our motivation is that the exact Green's function we derive provides a
valuable tool for examining nonperturbative fermion dynamics in classical
gluon backgrounds. Moreover, our result contributes to the wider program of
understanding the behavior of quantum fields in strong classical fields,
with implications for the physics of early-universe cosmology,
non-equilibrium quantum field theory, and non-Abelian kinetic theory.

The problem setting is presented in Section 2. Section 3 gives the
derivation of the Green's function. Section 4 is devoted to the properties
of the obtained Green's function and explores applications. The results are
summarized in Section 5. We list the four types of expressions that equal
zero in Appendix A. In Appendix B, we derive a relation between the
functions entering in the expression for the Green's function.

\section{Setting of the problem}

\label{problem}

In this article, we use the results obtained in \cite{Koshelkin2010}, where
the exact solution of the Dirac equation in the external Yang-Mills gauge
field was found. The fermion field $\psi \left( x\right) $ represents the
fundamental representations of the $SU(N)$ group, while the field $%
A_{a}^{\mu }\left( x\right) $ corresponds to the associated representation
of the $SU(N)$ group. The external non-Abelian field $A_{a}^{\mu }\left(
x\right) $ and the fermion field $\psi \left( x\right) $ satisfy the
Yang-Mills and Dirac equations, respectively.%
\begin{equation}
\left[ i\gamma ^{\mu }\left( \partial _{\mu }+igA_{\mu }^{a}\left( x\right)
T_{a}\right) -m\right] \psi \left( x\right) =0,  \label{Dirac eq}
\end{equation}%
\begin{equation}
\bar{\psi}\left( x\right) \left[ i\gamma ^{\mu }\left( \overleftarrow{%
\partial }_{\mu }-igA_{\mu }^{a}\left( x\right) T_{a}\right) +m\right] =0,\;%
\bar{\psi}\left( x\right) =\psi ^{\dag }\left( x\right) \gamma ^{0},
\label{Dirac coneq}
\end{equation}%
\begin{equation}
\partial _{\mu }F_{a}^{\nu \mu }\left( x\right) -gf_{ab}\,^{c}A_{\mu
}^{b}\left( x\right) F_{c}^{\nu \mu }\left( x\right) =0,
\label{Yang-Mills eq}
\end{equation}%
\begin{equation}
F_{a}^{\nu \mu }\left( x\right) =\partial ^{\nu }A_{a}^{\mu }\left( x\right)
-\partial ^{\mu }A_{a}^{\nu }\left( x\right) -gf_{a}\,^{bc}A_{b}^{\nu
}\left( x\right) A_{c}^{\mu }\left( x\right) ,  \label{Fanyumyu}
\end{equation}%
\ where we have made the following notations: $m$ is a fermion mass, $g$ is
the coupling constant, $\gamma ^{\mu }$ are the Dirac matrices, $x^{\mu
}=\left( x^{0},\vec{x}\right) $ is a vector in Minkowski spacetime, $%
\partial _{\mu }=\left( \partial /\partial t,\nabla \right) $. By the Roman
numerals, we enumerate the basis of the space of the associated
representation of the $SU(N)$ group, so that $a,b,c=1,\ldots ,\;N^{2}-1$. We
use the signature $G^{\mu \nu }=\mathrm{diag}\left( 1,-1,-1,-1\right) $ for
the metric tensor $G^{\mu \nu }$. It is implied that any pair of repeated
indexes must be summarized, and $T_{a}$ are the generators of the $SU\left(
N\right) $ group which the commutative relations $\left[ T_{a},T_{b}\right]
_{-}=T_{a}T_{b}-T_{b}T_{a}=if_{ab}\,^{c}T_{c}$\ and normalization condition $%
\mathrm{Tr}\left( T_{a}T_{b}\right) =\frac{1}{2}\delta _{ab},\;\left(
T_{a}\right) ^{\dagger }=T_{a}$. The real and anti-symmetric (with respect
to transposition in any pair of indices) $f_{ab}\,^{c}$ are the structure
constants of the $SU\left( N\right) $ group, and $\delta _{ab}$ is the
Kronecker delta. It is straightforward to verify that the generators of the
group $SU(N)$ satisfy the following conditions%
\begin{equation}
\left[ T_{a},T_{b}\right] _{+}=T_{a}T_{b}+T_{b}T_{a}=\frac{1}{N}\delta
_{ab}+d_{abc}T^{c},\;d_{abc}=\{Tr\left( \left[ T_{a},T_{b}\right]
_{+}T_{c}\right) ,  \label{generatoranticommut}
\end{equation}%
where $d_{abc}$ is real and symmetric under transposition of any pair of
indices.

As an external field, it was considered the solution of Eqs (\ref{Yang-Mills
eq}, \ref{Fanyumyu}) in the form of a plane wave on the light cone \cite%
{Coleman1977} as a one-component 4-vector in Minkowski space-time in the
axial gauge:%
\begin{equation}
A_{+}^{a}\left( x\right) =f^{a}\left( x^{+}\right) x^{1}+g^{a}\left(
x^{+}\right) x^{2}+h^{a}\left( x^{+}\right)
,\;A_{-}^{a}=A_{1}^{a}=A_{2}^{a}=0,  \label{Colemansolution}
\end{equation}
where $x^{\pm }=x^{0}\pm x^{3}$, $x^{1}$, and $x^{2}$ are the light-cone
coordinates, the $f^{a}\left( x^{+}\right) $, $g^{a}\left( x^{+}\right) $,
and $h^{a}\left( x^{+}\right) $ are arbitrary bounded functions. Note  that
the function $h^{a}\left( x^{+}\right) $ can be removed by a gauge
transformation. We have the following relations%
\begin{equation}
\partial ^{\mu }A_{\mu }^{a}\left( x\right) =k^{\mu }\dot{A}_{\mu
}^{a}\left( x\right) =\partial ^{+}A_{+}^{a}\left( x\right) =0,\;\varphi
=kx,\;k_{\mu }k^{\mu }=0\;\Rightarrow \;k^{\mu }A_{\mu }^{a}\left( x\right)
=0.  \label{kAortogonal}
\end{equation}%
The differentiation with respect to the variable $\varphi =kx$ is indicated
by the dot over the letter. It is easy to see that the non-Abelian field $%
A_{\mu }^{a}\left( \varphi \right) $ in the form of (\ref{Colemansolution})
satisfies the relation:%
\begin{equation}
A_{\mu }^{a}\left( \varphi \right) A_{b}^{\mu }\left( \varphi \right) =0,
\label{amyuamyuzero}
\end{equation}%
and Eq (\ref{Yang-Mills eq}).

On a shell where $p^{2}=m^{2}$, the general solution of the Dirac equation is%
\begin{align}
\psi \left( x\right) & =\sum_{\sigma ,\alpha }\int \frac{d^{3}p}{\sqrt{2p^{0}%
}\left( 2\pi \right) ^{3}}\left\{ \hat{a}_{\sigma ,\alpha }\left( p\right)
\psi _{\sigma ,\alpha }\left( x,p\right) +\hat{b}_{\sigma ,\alpha }^{\dag
}\left( p\right) \psi _{-\sigma ,\alpha }\left( x,-p\right) \right\} ,
\notag \\
\bar{\psi}\left( x\right) & =\sum_{\sigma ,\alpha }\int \frac{d^{3}p}{\sqrt{%
2p^{0}}\left( 2\pi \right) ^{3}}\left\{ \hat{a}_{\sigma ,\alpha }^{\dag
}\left( p\right) \bar{\psi}_{\sigma ,\alpha }\left( x,p\right) +\hat{b}%
_{\sigma ,\alpha }\left( p\right) \bar{\psi}_{-\sigma ,\alpha }\left(
x,-p\right) \right\} ,  \label{GenSolut}
\end{align}%
where
\begin{align}
\psi _{\sigma ,\alpha }\left( x,p\right) & =e^{-ipx}\cos \left( \theta
\right) \left\{ \left( 1-igT_{a}\frac{\tan \left( \theta \right) }{\theta
\left( pk\right) }\int_{0}^{\varphi }d\varphi ^{\prime }\left( A_{\mu
}^{a}p^{\mu }\right) \right) +\frac{g\left( \gamma ^{\mu }k_{\mu }\right)
\left( \gamma ^{\mu }A_{\mu }^{a}\right) }{2\left( pk\right) }\left[ \frac{%
\tan \left( \theta \right) }{\theta }T_{a}\right. \right.   \notag \\
& \left. \left. +\frac{g}{\left( pk\right) }\frac{1}{2N}\int_{0}^{\varphi
}d\varphi ^{\prime }\left( A_{\mu }^{a}p^{\mu }\right) \left( -i\frac{\tan
\left( \theta \right) }{\theta }+\frac{g}{\left( pk\right) }\frac{\theta
-\tan \left( \theta \right) }{\theta ^{3}}T_{b}\int_{0}^{\varphi }d\varphi
^{\prime }\left( A_{\mu }^{b}p^{\mu }\right) \right) \right] \right\}
\notag \\
& \times u_{\sigma }\left( p\right) v_{\alpha },  \notag \\
\theta & =\theta \left( p,\varphi \right) =\frac{g}{\left( pk\right) }\sqrt{%
\frac{1}{2N}}\left( \int_{0}^{\varphi }d\varphi ^{\prime }\left( A_{\mu
}^{a}\left( \varphi ^{\prime }\right) p^{\mu }\right) \int_{0}^{\varphi
}d\varphi ^{\prime \prime }\left( A_{a}^{\mu }\left( \varphi ^{\prime \prime
}\right) p_{\mu }\right) \right) ^{\frac{1}{2}},  \notag \\
p^{2}& =p^{\mu }p_{\mu }=m^{2}.  \label{exactsolut}
\end{align}%
where $u_{\sigma }\left( p\right) $ and $v_{\alpha }$ are some spinors which
are the elements of the spaces of the appropriate representations. Note that
the function $\psi _{\sigma ,\alpha }\left( x,p\right) $ depends on both the
spin variable $\sigma $ and the variable $\alpha $, where $\alpha $ = $1\div
N$, which describes a fermion's state in the space of the $SU\left( N\right)
$ group's fundamental representation. For them, the following relationships
occur:%
\begin{align}
\bar{u}_{\sigma }\left( p\right) u_{\lambda }\left( p^{\prime }\right) &
=\pm 2m\delta _{\sigma \lambda }\delta _{pp^{\prime
}},\;p^{2}=m^{2},\;v_{\alpha }^{\dagger }v_{\beta }=\delta _{\alpha \beta
},\;\mathrm{Tr}\left( T_{a}\right) =0,  \notag \\
u_{\sigma }\left( p\right) \bar{u}_{\lambda }\left( p\right) & =\left(
\gamma ^{\mu }p_{\mu }+m\right) _{\sigma \lambda },\;u_{\sigma }\left(
-p\right) \bar{u}_{\lambda }\left( -p\right) =\left( \gamma ^{\mu }p_{\mu
}-m\right) _{\sigma \lambda },  \label{conditions}
\end{align}%
where the symbols $\hat{a}_{\sigma ^{\prime },\alpha ^{\prime }}^{\dag
}\left( q\right) $; $\hat{b}_{\sigma ^{\prime },\alpha ^{\prime }}^{\dag
}\left( q\right) $ and $\hat{a}_{\sigma ,\alpha }\left( p\right) $; $\hat{b}%
_{\sigma ,\alpha }\left( p\right) $ are the operators of creation and
annihilation of a fermion ($\hat{a}_{\sigma ,\alpha }\left( p\right) $; $%
\hat{a}_{\sigma ^{\prime },\alpha ^{\prime }}^{\dag }\left( q\right) $) and
anti-fermion ($\hat{b}_{\sigma ,\alpha }\left( p\right) $; $\hat{b}_{\sigma
^{\prime },\alpha ^{\prime }}^{\dag }\left( q\right) $), correspondingly
\cite{Berestetzkii}. They satisfy the following relations for fermion
operators.
\begin{align}
\left[ \hat{a}_{\sigma ,\alpha }\left( p\right) ,\hat{a}_{\sigma ^{\prime
},\alpha ^{\prime }}^{\dag }\left( q\right) \right] _{+}& =\left( 2\pi
\right) ^{3}\delta \left( p-q\right) \delta _{\sigma \sigma ^{\prime
}}\delta _{\alpha \alpha ^{\prime }},  \notag \\
\left[ \hat{b}_{\sigma ,\alpha }\left( p\right) ,\hat{b}_{\sigma ^{\prime
},\alpha ^{\prime }}^{\dag }\left( q\right) \right] _{+}& =\left( 2\pi
\right) ^{3}\delta \left( p-q\right) \delta _{\sigma \sigma ^{\prime
}}\delta _{\alpha \alpha ^{\prime }}.  \label{operatorscommut}
\end{align}%
All the remaining anticommutators are equal zero. In addition, we have the
relations%
\begin{equation}
\langle 0|\hat{a}_{\sigma ,\alpha }\left( p\right) \hat{a}_{\sigma ,\alpha
}^{\dag }\left( p\right) |0\rangle =1,\;\langle 0|\hat{b}_{\sigma ,\alpha }%
\hat{b}_{\sigma ,\alpha }^{\dag }|0\rangle =1,  \label{vacuumrelations}
\end{equation}%
where the symbol $|0\rangle $ represents the state of the fermion vacuum.

The function (\ref{exactsolut}) is normalized by the condition:
\begin{equation}
\int dx^{3}\psi _{\sigma ,\alpha }^{\ast }\left( x,p^{\prime }\right) \psi
_{\sigma ,\alpha }\left( x,p\right) =\left( 2\pi \right) ^{3}\delta \left(
p-p^{\prime }\right) .  \label{normcondition}
\end{equation}%
In the next section, we calculate the Green's functions of fermions in the
presence of an external Yang-Mills gauge field.

\section{Green's function of fermions in an external Yang-Mills gauge field}

\label{Green}

Let us examine the fermionic Green's function. According to \cite%
{Berestetzkii}, we can write for the Green's function:%
\begin{equation}
G_{F}\left( x,x^{\prime }\right) =-i\langle 0|\psi \left( x\right) \bar{\psi}%
\left( x^{\prime }\right) |0\rangle \Theta \left( x^{0}-x^{0\prime }\right)
+i\langle 0|\bar{\psi}\left( x^{\prime }\right) \psi \left( x\right)
|0\rangle \Theta \left( x^{0\prime }-x^{0}\right) ,  \label{Greendefin}
\end{equation}%
where $\Theta \left( x\right) $ is the unit step function.

Substituting (\ref{GenSolut}) and (\ref{exactsolut}) into (\ref{Greendefin})
and utilizing (\ref{conditions}), (\ref{operatorscommut}), and (\ref%
{vacuumrelations}), while taking into account that for a homogeneous and
isotropic fermion system and condition (\ref{kAortogonal}), certain types of
expressions are equal to zero due to relativistic invariance (see Appendix
A; all these expressions take the forms of the four types discussed in
Appendix A), we derive the following expression for Green's function:%
\begin{align}
G_{F}\left( x,x^{\prime }\right) & =-iN\int \frac{d^{3}p}{\left( 2\pi
\right) ^{3}}\left[ \frac{1}{2p^{0}}\left( \gamma ^{\mu }p_{\mu }+m\right)
e^{-ip\left( x-x^{\prime }\right) }U\left( p\right) \Theta \left(
x^{0}-x^{0\prime }\right) \right.  \notag \\
& \left. -\frac{1}{2p^{0}}\left( \gamma ^{\mu }p_{\mu }-m\right)
e^{-ip\left( x^{\prime }-x\right) }W\left( p\right) \Theta \left( x^{0\prime
}-x^{0}\right) \right] ,  \label{Green1}
\end{align}%
where%
\begin{align}
U\left( p\right) & =\cos \left( \theta \left( p,\varphi \right) \right) \cos
\left( \theta \left( p,\varphi ^{\prime }\right) \right) \left\{ 1+\frac{%
\tan \left( \theta \left( p,\varphi ^{\prime }\right) \right) }{\theta
\left( p,\varphi ^{\prime }\right) }\frac{g\left( \left( \gamma ^{\sigma
}\right) ^{\dagger }A_{\sigma }^{e}\left( \varphi ^{\prime }\right) \right)
\left( \left( \gamma ^{\rho }\right) ^{\dagger }k_{\rho }\right) }{2\left(
pk\right) }T_{e}\right.  \notag \\
& \left. +\frac{g\left( \gamma ^{\nu }k_{\nu }\right) \left( \gamma
^{\lambda }A_{\lambda }^{e}\left( \varphi \right) \right) }{2\left(
pk\right) }\frac{\tan \left( \theta \left( p,\varphi \right) \right) }{%
\theta \left( p,\varphi \right) }T_{e}\right.  \notag \\
& \left. +\frac{g\left( \gamma ^{\nu }k_{\nu }\right) \left( \gamma
^{\lambda }A_{\lambda }^{b}\left( \varphi \right) \right) }{2\left(
pk\right) }\frac{\tan \left( \theta \left( p,\varphi \right) \right) }{%
\theta \left( p,\varphi \right) }\frac{\tan \left( \theta \left( p,\varphi
^{\prime }\right) \right) }{\theta \left( p,\varphi ^{\prime }\right) }\frac{%
g\left( \left( \gamma ^{\sigma }\right) ^{\dagger }A_{\sigma }^{e}\left(
\varphi ^{\prime }\right) \right) \left( \left( \gamma ^{\rho }\right)
^{\dagger }k_{\rho }\right) }{2\left( pk\right) }T_{b}T_{e}\right\} ,
\label{Udefin}
\end{align}%
\begin{align}
W\left( p\right) & =\cos \left( \theta \left( p,\varphi \right) \right) \cos
\left( \theta \left( p,\varphi ^{\prime }\right) \right) \left\{ 1-\frac{%
g\left( \gamma ^{\nu }k_{\nu }\right) \left( \gamma ^{\lambda }A_{\lambda
}^{e}\left( \varphi \right) \right) }{2\left( pk\right) }\frac{\tan \left(
\theta \left( p,\varphi \right) \right) }{\theta \left( p,\varphi \right) }%
T_{e}\right.  \notag \\
& \left. -\frac{\tan \left( \theta \left( p,\varphi ^{\prime }\right)
\right) }{\theta \left( p,\varphi ^{\prime }\right) }\frac{g\left( \left(
\gamma ^{\sigma }\right) ^{\dagger }A_{\sigma }^{e}\left( \varphi ^{\prime
}\right) \right) \left( \left( \gamma ^{\rho }\right) ^{\dagger }k_{\rho
}\right) }{2\left( pk\right) }T_{e}\right.  \notag \\
& \left. +\frac{\tan \left( \theta \left( p,\varphi ^{\prime }\right)
\right) }{\theta \left( p,\varphi ^{\prime }\right) }\frac{g\left( \left(
\gamma ^{\sigma }\right) ^{\dagger }A_{\sigma }^{e}\left( \varphi ^{\prime
}\right) \right) \left( \left( \gamma ^{\rho }\right) ^{\dagger }k_{\rho
}\right) }{2\left( pk\right) }\frac{g\left( \gamma ^{\nu }k_{\nu }\right)
\left( \gamma ^{\lambda }A_{\lambda }^{b}\left( \varphi \right) \right) }{%
2\left( pk\right) }\frac{\tan \left( \theta \left( p,\varphi \right) \right)
}{\theta \left( p,\varphi \right) }T_{e}T_{b}\right\} ,  \label{Wdefin}
\end{align}%
with $\varphi =kx$ and $\varphi ^{\prime }=kx^{\prime }$. In the second term
of (\ref{Green1}), relabeling $\vec{p}$ by $-\vec{p}$, we can write%
\begin{align}
G_{F}\left( x,x^{\prime }\right) & =-iN\int \frac{d^{3}p}{\left( 2\pi
\right) ^{3}}e^{i\vec{p}\left( \vec{x}-\vec{x}^{\prime }\right) }\left[
\frac{1}{2p^{0}}\left( \gamma ^{\mu }p_{\mu }+m\right) e^{-ip\left(
x^{0}-x^{0\prime }\right) }U\left( p\right) \Theta \left( x^{0}-x^{0\prime
}\right) \right.  \notag \\
& \left. -\frac{1}{2p^{0}}\left( \gamma ^{\mu }p_{\mu }-m\right)
e^{-ip\left( x^{0\prime }-x^{0}\right) }W\left( p^{0},-\vec{p}\right) \Theta
\left( x^{0\prime }-x^{0}\right) \right] .  \label{Green2}
\end{align}

For convenience, we will rewrite the formula in the following form%
\begin{align}
G_{F}\left( x,x^{\prime }\right) & =-iN\int \frac{d^{3}p}{\left( 2\pi
\right) ^{3}}e^{i\vec{p}\left( \vec{x}-\vec{x}^{\prime }\right) }\left[
\frac{1}{2E_{p}}\left( \gamma ^{\mu }p_{\mu }+m\right) e^{-ip\left(
x^{0}-x^{0\prime }\right) }U\left( p\right) \Theta \left( x^{0}-x^{0\prime
}\right) \right.  \notag \\
& \left. -\frac{1}{2E_{p}}\left( \gamma ^{\mu }p_{\mu }-m\right)
e^{-ip\left( x^{0\prime }-x^{0}\right) }W\left( E_{p},-\vec{p}\right) \Theta
\left( x^{0\prime }-x^{0}\right) \right] ,  \label{Green3}
\end{align}%
where $\left( p^{0},\vec{p}\right) =\left( E_{p},\vec{p}\right) $.

We rewrite the term in brackets using Cauchy's integral formula.

For a holomorphic function $g\left( z\right) $ defined in an open subset $Q$
of $\mathcal{C}$, consider a closed disk $D$ with $C=\partial D\subset Q$.
From Cauchy's formula one has%
\begin{equation}
g\left( z_{0}\right) =\frac{1}{2\pi i}\int_{C}\frac{g\left( z\right) }{%
z-z_{0}}dz.  \label{Cauchy}
\end{equation}%
Using this formula, the first term in the (\ref{Green3}) can be expressed as%
\begin{align}
& \frac{1}{2E_{p}}\left( \gamma ^{\mu }p_{\mu }+m\right) e^{-ip\left(
x^{0}-x^{0\prime }\right) }U\left( p\right) \Theta \left( x^{0}-x^{0\prime
}\right)  \notag \\
& =-\frac{\Theta \left( x^{0}-x^{0\prime }\right) }{2\pi i}\int_{C_{1}}dp^{0}%
\frac{\left( \gamma ^{\mu }p_{\mu }+m\right) e^{-ip^{0}\left(
x^{0}-x^{0\prime }\right) }}{\left( p^{0}-E_{p}\right) \left(
p^{0}+E_{p}\right) }U\left( p\right) ,  \label{Greenp01}
\end{align}%
where the poles should be avoided in the $\epsilon $-surroundings as drawn
in the Fig.1, left panel, and the contour is closed in the lower half-plane,
such as to pick up the residue at $p^{0}=+E_{p}$, and the integral is
clockwise, which clarifies the overall minus sign. Note that this result is
true for any contour that encloses the poles as drawn. Since $%
x^{0}-x^{0\prime }>0$, the integral along the lower half-plane
asymptotically vanishes if we choose to deform the contour to infinity,
corresponding to $R\rightarrow \infty $. For the second term in the (\ref%
{Green3}), we have%
\begin{align}
& -\frac{1}{2E_{p}}\left( \gamma ^{\mu }p_{\mu }-m\right) e^{ip\left(
x^{0}-x^{0\prime }\right) }W\left( p^{0},-\vec{p}\right) \Theta \left(
x^{0\prime }-x^{0}\right)  \notag \\
& =\Theta \left( x^{0\prime }-x^{0}\right) \frac{1}{2\pi i}\int_{C_{2}}dp^{0}%
\frac{\left( \gamma ^{\mu }p_{\mu }-m\right) e^{+ip^{0}\left(
x^{0}-x^{0\prime }\right) }}{\left( p^{0}-E_{p}\right) \left(
p^{0}+E_{p}\right) }W\left( p^{0},-\vec{p}\right) ,  \label{Greenp02}
\end{align}%
and $C_{2}$ runs counter-clockwise this time (see Fig.1 right
panel), but it catches the pole at $p^{0}=-E_{p}$, which results in
an overall minus. Both expressions hold for any $R>E_{p}$, but if
$R\rightarrow \infty $, then the integral in the lower/upper
half-plane each vanishes due to the appearance of $\Theta \left(
x^{0}-x^{0\prime }\right) $ and $\Theta \left( x^{0\prime
}-x^{0}\right) $, respectively. The order of time becomes crucial at
this point.
\begin{figure}[tbph]
\begin{center}
\begin{tabular}{cc}
\epsfig{figure=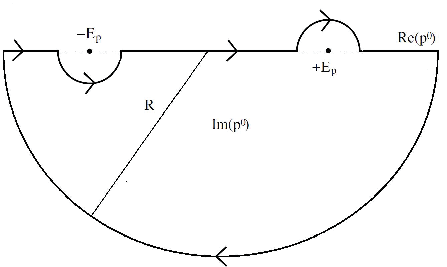,width=7.cm,height=4.cm} & \quad %
\epsfig{figure=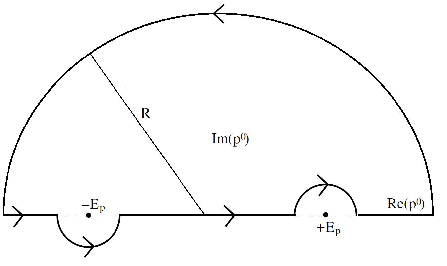,width=7.cm,height=4.cm}%
\end{tabular}%
\end{center}
\caption{Integration contours C$_{1}$ (left panel) and C$_{2}$ (right
panel). }
\end{figure}

By substituting the expressions (\ref{Greenp01}) and (\ref{Greenp02}) in (%
\ref{Green3}), we will have%
\begin{align}
G_{F}\left( x,x^{\prime }\right) & =N\int \frac{d^{3}p}{\left( 2\pi \right)
^{4}}\left[ \Theta \left( x^{0}-x^{0\prime }\right) \int_{C_{1}}dp^{0}\frac{%
\left( \gamma ^{\mu }p_{\mu }+m\right) e^{-ip\left( x-x^{\prime }\right) }}{%
p^{2}-m^{2}}U\left( p^{0},\vec{p}\right) \right.  \notag \\
& \left. -\Theta \left( x^{0\prime }-x^{0}\right) \int_{C_{2}}dp^{0}\frac{%
\left( \gamma ^{\mu }p_{\mu }-m\right) e^{ip\left( x-x^{\prime }\right) }}{%
p^{2}-m^{2}}W\left( p^{0},+\vec{p}\right) \right] .  \label{Greenfinala}
\end{align}%
It is easy to observe that $W\left( -p\right) =U\left( p\right) $ (see
Appendix B) and using that $\Theta \left( x^{0}-x^{0\prime }\right) +\Theta
\left( x^{0\prime }-x^{0}\right) =1$ and
\begin{align}
& \left( p^{0}-E_{p}+i\frac{\tilde{\epsilon}}{E_{p}}\right) \left(
p^{0}+E_{p}-i\frac{\tilde{\epsilon}}{E_{p}}\right)  \notag \\
& =\left( p^{0}\right) ^{2}-E_{p}^{2}+2i\tilde{\epsilon}+\frac{\tilde{%
\epsilon}^{2}}{E_{p}^{2}}=p^{2}-m^{2}+i\epsilon ,\;\epsilon =2\tilde{\epsilon%
}-\frac{\tilde{\epsilon}}{E_{p}},  \label{some transformations}
\end{align}%
we have%
\begin{equation}
G_{F}\left( x,x^{\prime }\right) =N\int_{C}\frac{d^{4}p}{\left( 2\pi \right)
^{4}}\frac{\left( \gamma ^{\mu }p_{\mu }+m\right) U\left( p\right) }{%
p^{2}-m^{2}+i\epsilon }e^{-ip\left( x-x^{\prime }\right) }.
\label{Greenfunction}
\end{equation}%
where the contour $C$ is drawn in Fig.2 on the left panel. Since this is a
complex contour integral, all that matters is the relative placement of the
contour to the poles. Therefore, we can equivalently pick the contour $C$ on
top of the real axis but shift the poles, e.g., by an amount $\pm i\tilde{%
\epsilon}/E_{p}$ in the limit $\tilde{\epsilon}\rightarrow 0$ (see Fig.2
right panel).
\begin{figure}[tbph]
\begin{center}
\begin{tabular}{cc}
\epsfig{figure=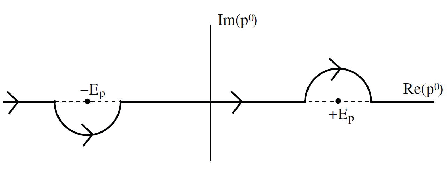,width=7.cm,height=3.cm} & \quad %
\epsfig{figure=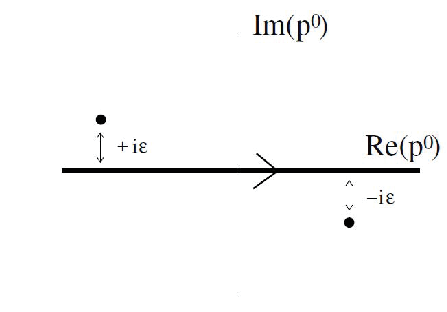,width=7.cm,height=3.cm}%
\end{tabular}%
\end{center}
\caption{Contour C for the Feynman propagator on the left panel and shifted
poles on the right panel.}
\end{figure}

For $\tilde{G}_{F}\left( x,x^{\prime }\right) =\langle 0|\psi \left(
x\right) \bar{\psi}\left( x^{\prime }\right) |0\rangle \Theta \left(
t-t^{\prime }\right) -\langle 0|\bar{\psi}\left( x^{\prime }\right) \psi
\left( x\right) |0\rangle \Theta \left( t^{\prime }-t\right) $ we have the
following expression%
\begin{equation}
\tilde{G}_{F}\left( x,x^{\prime }\right) =N\int_{C}\frac{d^{4}p}{\left( 2\pi
\right) ^{4}}\frac{i\left( \gamma ^{\mu }p_{\mu }+m\right) U\left( p\right)
}{p^{2}-m^{2}+i\epsilon }e^{-ip\left( x-x^{\prime }\right) }.
\label{Greenfunction1}
\end{equation}%
When $A_{\mu }^{a}\left( x\right) =0$, it follows that $U\left( p\right) =1$%
, and setting $N=1$ yields the free fermion propagator:%
\begin{equation}
\tilde{G}_{F,A=0}\left( x,x^{\prime }\right) =\int_{C}\frac{d^{4}p}{\left(
2\pi \right) ^{4}}\frac{i\left( \gamma ^{\mu }p_{\mu }+m\right) }{%
p^{2}-m^{2}+i\epsilon }e^{-ip\left( x-x^{\prime }\right) }.
\label{Greenfunctionfree}
\end{equation}

\section{Properties of the Green's function and its applications}

\label{propappl}

Among the properties of Green's function (\ref{Greenfunction}), the
following should be noted: 1. Green's function transforms covariantly under
local gauge transformations of the external field. 2. In the fundamental
representation of $SU(N)$ the internal symmetry generators $T_{a}$ appear
explicitly in the expansion of $U(p)$. 3. The propagator takes into account
the infinite series of interactions between the fermion and the external
field (via the $U(p)$ factor) rather than truncating perturbatively. 4. The
gamma matrices and the presence of conjugated gamma structures reflect
spin-dependent interactions and helicity transitions induced by the
background. 5. The integration over $p$ and the appearance of $%
p^{2}-m^{2}+i\epsilon $\ indicate causal Feynman propagation with dressed
vertices encoded in $U\left( p\right) $. 6. The color interference and
memory effects typical of non-Abelian dynamics in the higher-order terms are
described by the structure $T_{e}T_{b}$.

We pay regard to the following utilizations of the obtained Green's
function: 1. Useful in modeling fermion propagation through strong gluonic
backgrounds in heavy-ion collisions or dense nuclear matter. 2. For the
study of quark-gluon plasma evolution and parton saturation phenomena, the
Green function is relevant. 3. Used in computing $S$-matrix elements
involving external fields (e.g., non-Abelian Compton scattering, vacuum
birefringence, etc.). 4. Forms the basis of approximation methods where the
external field is classical, but fermion dynamics is treated fully quantum
mechanically.

\section{Conclusion}

\label{Concl}

We derived the Green's function for fermions in an external Yang-Mills gauge
field with an $SU(N)$ symmetry group. The external field is analyzed as a
plane-wave solution to the Yang-Mills equation, propagating along the light
cone. Some properties and applications of the Green's function have been
outlined. The color precession will be discussed in future work.

\appendix

\section{Vanishing Integrals: Four Classes of Zero-Valued Contributions}

\label{AppA}

We describe the four types of expressions that equal zero.

\begin{align}
& i\int \frac{d^{3}pN}{2p^{0}\left( 2\pi \right) ^{3}}\left( \gamma ^{\mu
}p_{\mu }+m\right) e^{-ip\left( x-x^{\prime }\right) }\cos \left( \theta
\left( p,\varphi \right) \right) \cos \left( \theta \left( p,\varphi
^{\prime }\right) \right)  \notag \\
& \times T_{a}T_{d}g^{2}\frac{\tan \left( \theta \left( p,\varphi \right)
\right) }{\theta \left( p,\varphi \right) \left( pk\right) }\frac{\tan
\left( \theta \left( p,\varphi ^{\prime }\right) \right) }{\theta \left(
p,\varphi ^{\prime }\right) \left( pk\right) }\int_{0}^{\varphi }d\varphi
^{\prime }\left( A_{\mu }^{a}\left( \varphi ^{\prime }\right) p^{\mu
}\right) \int_{0}^{\varphi ^{\prime }}d\varphi ^{\prime \prime }\left(
A_{\gamma }^{d}\left( \varphi ^{\prime \prime }\right) p^{\gamma }\right)
\notag \\
& =iNT_{a}T_{d}g^{2}\int_{0}^{\varphi }d\varphi ^{\prime }A_{\alpha
}^{a}\left( \varphi ^{\prime }\right) \int_{0}^{\varphi ^{\prime }}d\varphi
^{\prime \prime }A_{\gamma }^{d}\left( \varphi ^{\prime \prime }\right) \int
\frac{d^{3}p}{2p^{0}\left( 2\pi \right) ^{3}}\left( \gamma ^{\mu }p_{\mu
}+m\right)  \notag \\
& \times e^{-ip\left( x-x^{\prime }\right) }\cos \left( \theta \left(
p,\varphi \right) \right) \cos \left( \theta \left( p,\varphi ^{\prime
}\right) \right) \frac{\tan \left( \theta \left( p,\varphi \right) \right) }{%
\theta \left( p,\varphi \right) \left( pk\right) }\frac{\tan \left( \theta
\left( p,\varphi ^{\prime }\right) \right) }{\theta \left( p,\varphi
^{\prime }\right) \left( pk\right) }p^{\alpha }p^{\gamma }  \notag \\
& =iNT_{a}T_{d}g^{2}\int_{0}^{\varphi }d\varphi ^{\prime }A_{\alpha
}^{a}\left( \varphi ^{\prime }\right) \int_{0}^{\varphi ^{\prime }}d\varphi
^{\prime \prime }A_{\gamma }^{d}\left( \varphi ^{\prime \prime }\right)
k^{\alpha }k^{\gamma }f_{1}\left( \varphi ,\varphi ^{\prime }\right) =0,
\label{app1}
\end{align}%
\begin{align}
& i\int \frac{d^{3}pN}{2p^{0}\left( 2\pi \right) ^{3}}\left( \gamma ^{\mu
}p_{\mu }+m\right) e^{-ip\left( x-x^{\prime }\right) }\cos \left( \theta
\left( p,\varphi \right) \right) \cos \left( \theta \left( p,\varphi
^{\prime }\right) \right)  \notag \\
& \times \frac{g}{\left( pk\right) }\frac{1}{2N}\int_{0}^{\varphi ^{\prime
}}d\varphi ^{\prime \prime }\left( A_{e\delta }\left( \varphi ^{\prime
\prime }\right) p^{\delta }\right) \left( i\frac{\tan \left( \theta \left(
p,\varphi ^{\prime }\right) \right) }{\theta \left( p,\varphi ^{\prime
}\right) }\right.  \notag \\
& \left. +\frac{g}{\left( pk\right) }\frac{\theta \left( p,\varphi ^{\prime
}\right) -\tan \left( \theta \left( p,\varphi ^{\prime }\right) \right) }{%
\theta ^{3}\left( p,\varphi ^{\prime }\right) }T_{f}\int_{0}^{\varphi
^{\prime }}d\varphi ^{\prime \prime }\left( A_{\eta }^{f}\left( \varphi
^{\prime \prime }\right) p^{\eta }\right) \right) \frac{g\left( \left(
\gamma ^{\sigma }\right) ^{\dagger }A_{\sigma }^{e}\left( \varphi ^{\prime
}\right) \right) \left( \left( \gamma ^{\rho }\right) ^{\dagger }k_{\rho
}\right) }{2\left( pk\right) }  \notag \\
& =\frac{i}{2}\int_{0}^{\varphi ^{\prime }}d\varphi ^{\prime \prime
}A_{e\delta }\left( \varphi ^{\prime \prime }\right) \int \frac{%
d^{3}pp^{\delta }}{2p^{0}\left( 2\pi \right) ^{3}}\frac{g}{\left( pk\right) }%
\left( \gamma ^{\mu }p_{\mu }+m\right) e^{-ip\left( x-x^{\prime }\right)
}\cos \left( \theta \left( p,\varphi \right) \right) \cos \left( \theta
\left( p,\varphi ^{\prime }\right) \right)  \notag \\
& \times i\frac{\tan \left( \theta \left( p,\varphi ^{\prime }\right)
\right) }{\theta \left( p,\varphi ^{\prime }\right) }\frac{g\left( \left(
\gamma ^{\sigma }\right) ^{\dagger }A_{\sigma }^{e}\left( \varphi ^{\prime
}\right) \right) \left( \left( \gamma ^{\rho }\right) ^{\dagger }k_{\rho
}\right) }{2\left( pk\right) }  \notag \\
& +\frac{i}{2}\int_{0}^{\varphi ^{\prime }}d\varphi ^{\prime \prime
}A_{e\delta }\left( \varphi ^{\prime \prime }\right) \int_{0}^{\varphi
^{\prime }}d\varphi ^{\prime \prime }A_{\eta }^{f}\left( \varphi ^{\prime
\prime }\right) \int \frac{d^{3}pp^{\delta }p^{\eta }}{2p^{0}\left( 2\pi
\right) ^{3}}\frac{g}{\left( pk\right) }\left( \gamma ^{\mu }p_{\mu
}+m\right) e^{-ip\left( x-x^{\prime }\right) }  \notag \\
& \times \cos \left( \theta \left( p,\varphi \right) \right) \cos \left(
\theta \left( p,\varphi ^{\prime }\right) \right) \frac{g}{\left( pk\right) }%
\frac{\theta \left( p,\varphi ^{\prime }\right) -\tan \left( \theta \left(
p,\varphi ^{\prime }\right) \right) }{\theta ^{3}\left( p,\varphi ^{\prime
}\right) }T_{f}\frac{g\left( \left( \gamma ^{\sigma }\right) ^{\dagger
}A_{\sigma }^{e}\left( \varphi ^{\prime }\right) \right) \left( \left(
\gamma ^{\rho }\right) ^{\dagger }k_{\rho }\right) }{2\left( pk\right) }
\notag \\
& =\frac{i}{2}\int_{0}^{\varphi ^{\prime }}d\varphi ^{\prime \prime
}A_{e\delta }\left( \varphi ^{\prime \prime }\right) k^{\delta
}f_{2}^{e}\left( \varphi ^{\prime }\right)  \notag \\
& +\frac{i}{2}T_{f}\int_{0}^{\varphi ^{\prime }}d\varphi ^{\prime \prime
}A_{e\delta }\left( \varphi ^{\prime \prime }\right) \int_{0}^{\varphi
^{\prime }}d\varphi ^{\prime \prime }A_{\eta }^{f}\left( \varphi ^{\prime
\prime }\right) k^{\delta }k^{\eta }f_{3}^{e}\left( \varphi ^{\prime
}\right) =0,  \label{app2}
\end{align}

\begin{align}
& i\int \frac{d^{3}pN}{2p^{0}\left( 2\pi \right) ^{3}}\left( \gamma ^{\mu
}p_{\mu }+m\right) e^{-ip\left( x-x^{\prime }\right) }\cos \left( \theta
\left( p,\varphi \right) \right) \cos \left( \theta \left( p,\varphi
^{\prime }\right) \right)  \notag \\
& \times \left( -igT_{a}\frac{\tan \left( \theta \left( p,\varphi \right)
\right) }{\theta \left( p,\varphi \right) \left( pk\right) }%
\int_{0}^{\varphi }d\varphi ^{\prime }\left( A_{\mu }^{a}\left( \varphi
^{\prime }\right) p^{\mu }\right) \frac{\tan \left( \theta \left( p,\varphi
^{\prime }\right) \right) }{\theta \left( p,\varphi ^{\prime }\right) }T_{e}%
\frac{g\left( \left( \gamma ^{\sigma }\right) ^{\dagger }A_{\sigma
}^{e}\left( \varphi ^{\prime }\right) \right) \left( \left( \gamma ^{\rho
}\right) ^{\dagger }k_{\rho }\right) }{2\left( pk\right) }\right)  \notag \\
& =Ng\int_{0}^{\varphi }d\varphi ^{\prime }A_{\alpha }^{a}\left( \varphi
^{\prime }\right) \int \frac{d^{3}pp^{\alpha }}{2p^{0}\left( 2\pi \right)
^{3}}\left( \gamma ^{\mu }p_{\mu }+m\right) e^{-ip\left( x-x^{\prime
}\right) }\cos \left( \theta \left( p,\varphi \right) \right) \cos \left(
\theta \left( p,\varphi ^{\prime }\right) \right)  \notag \\
& \times T_{a}\frac{\tan \left( \theta \left( p,\varphi \right) \right) }{%
\theta \left( p,\varphi \right) \left( pk\right) }\frac{\tan \left( \theta
\left( p,\varphi ^{\prime }\right) \right) }{\theta \left( p,\varphi
^{\prime }\right) }T_{e}\frac{g\left( \left( \gamma ^{\sigma }\right)
^{\dagger }A_{\sigma }^{e}\left( \varphi ^{\prime }\right) \right) \left(
\left( \gamma ^{\rho }\right) ^{\dagger }k_{\rho }\right) }{2\left(
pk\right) }  \notag \\
& =Ng\int_{0}^{\varphi }d\varphi ^{\prime }A_{\alpha }^{a}\left( \varphi
^{\prime }\right) k^{\alpha }T_{a}T_{e}f_{4}^{e}\left( \varphi ,\varphi
^{\prime }\right) =0,  \label{app3}
\end{align}%
\begin{align}
& \int \frac{d^{3}pN}{2p^{0}\left( 2\pi \right) ^{3}}\left( \gamma ^{\mu
}p_{\mu }+m\right) e^{-ip\left( x-x^{\prime }\right) }\cos \left( \theta
\left( p,\varphi \right) \right) \cos \left( \theta \left( p,\varphi
^{\prime }\right) \right)  \notag \\
& \times gT_{a}\frac{\tan \left( \theta \left( p,\varphi \right) \right) }{%
\theta \left( p,\varphi \right) \left( pk\right) }\int_{0}^{\varphi
}d\varphi ^{\prime }\left( A_{\mu }^{a}\left( \varphi ^{\prime }\right)
p^{\mu }\right) \frac{g}{\left( pk\right) }\frac{1}{2N}\int_{0}^{\varphi
^{\prime }}d\varphi ^{\prime \prime }\left( A_{e\delta }\left( \varphi
^{\prime \prime }\right) p^{\delta }\right)  \notag \\
& \times \left( i\frac{\tan \left( \theta \left( p,\varphi ^{\prime }\right)
\right) }{\theta \left( p,\varphi ^{\prime }\right) }+\frac{g}{\left(
pk\right) }\frac{\theta \left( p,\varphi ^{\prime }\right) -\tan \left(
\theta \left( p,\varphi ^{\prime }\right) \right) }{\theta ^{3}\left(
p,\varphi ^{\prime }\right) }T_{f}\int_{0}^{\varphi ^{\prime }}d\varphi
^{\prime \prime }\left( A_{\eta }^{f}\left( \varphi ^{\prime \prime }\right)
p^{\eta }\right) \right)  \notag \\
& \times \frac{g\left( \left( \gamma ^{\sigma }\right) ^{\dagger }A_{\sigma
}^{e}\left( \varphi ^{\prime }\right) \right) \left( \left( \gamma ^{\rho
}\right) ^{\dagger }k_{\rho }\right) }{2\left( pk\right) }  \notag \\
& =\frac{i}{2}\int_{0}^{\varphi }d\varphi ^{\prime }A_{\alpha }^{a}\left(
\varphi ^{\prime }\right) \int_{0}^{\varphi ^{\prime }}d\varphi ^{\prime
\prime }A_{e\delta }\left( \varphi ^{\prime \prime }\right) \int \frac{%
d^{3}pp^{\alpha }p^{\delta }}{2p^{0}\left( 2\pi \right) ^{3}}\left( \gamma
^{\mu }p_{\mu }+m\right) e^{-ip\left( x-x^{\prime }\right) }  \notag \\
& \times \cos \left( \theta \left( p,\varphi \right) \right) \cos \left(
\theta \left( p,\varphi ^{\prime }\right) \right) gT_{a}\frac{\tan \left(
\theta \left( p,\varphi \right) \right) }{\theta \left( p,\varphi \right)
\left( pk\right) }\frac{\tan \left( \theta \left( p,\varphi ^{\prime
}\right) \right) }{\theta \left( p,\varphi ^{\prime }\right) }\frac{g\left(
\left( \gamma ^{\sigma }\right) ^{\dagger }A_{\sigma }^{e}\left( \varphi
^{\prime }\right) \right) \left( \left( \gamma ^{\rho }\right) ^{\dagger
}k_{\rho }\right) }{2\left( pk\right) }  \notag \\
& +\frac{1}{2}\int_{0}^{\varphi }d\varphi ^{\prime }A_{\alpha }^{a}\left(
\varphi ^{\prime }\right) \int_{0}^{\varphi ^{\prime }}d\varphi ^{\prime
\prime }A_{e\delta }\left( \varphi ^{\prime \prime }\right)
\int_{0}^{\varphi ^{\prime }}d\varphi ^{\prime \prime }A_{\eta }^{f}\left(
\varphi ^{\prime \prime }\right)  \notag \\
& \times \int \frac{d^{3}pp^{\alpha }p^{\delta }p^{\eta }}{2p^{0}\left( 2\pi
\right) ^{3}}\left( \gamma ^{\mu }p_{\mu }+m\right) e^{-ip\left( x-x^{\prime
}\right) }\cos \left( \theta \left( p,\varphi \right) \right) \cos \left(
\theta \left( p,\varphi ^{\prime }\right) \right)  \notag \\
& \times gT_{a}\frac{\tan \left( \theta \left( p,\varphi \right) \right) }{%
\theta \left( p,\varphi \right) \left( pk\right) }\frac{g}{\left( pk\right) }%
\frac{\theta \left( p,\varphi ^{\prime }\right) -\tan \left( \theta \left(
p,\varphi ^{\prime }\right) \right) }{\theta ^{3}\left( p,\varphi ^{\prime
}\right) }T_{f}\frac{g\left( \left( \gamma ^{\sigma }\right) ^{\dagger
}A_{\sigma }^{e}\left( \varphi ^{\prime }\right) \right) \left( \left(
\gamma ^{\rho }\right) ^{\dagger }k_{\rho }\right) }{2\left( pk\right) }
\notag \\
& =\frac{i}{2}g\int_{0}^{\varphi }d\varphi ^{\prime }A_{\alpha }^{a}\left(
\varphi ^{\prime }\right) \int_{0}^{\varphi ^{\prime }}d\varphi ^{\prime
\prime }A_{e\delta }\left( \varphi ^{\prime \prime }\right) k^{\alpha
}k^{\delta }T_{a}f_{5}^{e}\left( \varphi ,\varphi ^{\prime }\right)  \notag
\\
& +\frac{1}{2}g\int_{0}^{\varphi }d\varphi ^{\prime }A_{\alpha }^{a}\left(
\varphi ^{\prime }\right) \int_{0}^{\varphi ^{\prime }}d\varphi ^{\prime
\prime }A_{e\delta }\left( \varphi ^{\prime \prime }\right)
\int_{0}^{\varphi ^{\prime }}d\varphi ^{\prime \prime }A_{\eta }^{f}\left(
\varphi ^{\prime \prime }\right) T_{a}T_{f}  \notag \\
& \times \left( k^{\alpha }k^{\delta }k^{\eta }f_{6}^{af}\left( \varphi
,\varphi ^{\prime }\right) +k^{\alpha }G^{\delta \eta }f_{7}^{af}\left(
\varphi ,\varphi ^{\prime }\right) \right.  \notag \\
& \left. +k^{\delta }G^{\alpha \eta }f_{8}^{af}\left( \varphi ,\varphi
^{\prime }\right) +k^{\eta }G^{\alpha \delta }f_{9}^{af}\left( \varphi
,\varphi ^{\prime }\right) \right) =0.  \label{app4}
\end{align}%
Here, $f_{i}$ are some functions.

\section{Proof of a Identity Involving Dirac Gamma Matrices}

\label{ApplB}

To prove that $W(-p)=U(p)$ , we need to verify that%
\begin{align}
& \left( \left( \gamma ^{\sigma }\right) ^{\dagger }A_{\sigma }^{e}\left(
\varphi ^{\prime }\right) \right) \left( \left( \gamma ^{\rho }\right)
^{\dagger }k_{\rho }\right) \left( \gamma ^{\nu }k_{\nu }\right) \left(
\gamma ^{\lambda }A_{\lambda }^{b}\left( \varphi \right) \right)  \notag \\
& =\left( \gamma ^{\nu }k_{\nu }\right) \left( \gamma ^{\lambda }A_{\lambda
}^{b}\left( \varphi \right) \right) \left( \left( \gamma ^{\sigma }\right)
^{\dagger }A_{\sigma }^{e}\left( \varphi ^{\prime }\right) \right) \left(
\left( \gamma ^{\rho }\right) ^{\dagger }k_{\rho }\right) .  \label{B1}
\end{align}%
It is enough to prove that%
\begin{equation}
\left( \gamma ^{\sigma }\right) ^{\dagger }\left( \gamma ^{\rho }\right)
^{\dagger }\gamma ^{\nu }\gamma ^{\lambda }=\gamma ^{\nu }\gamma ^{\lambda
}\left( \gamma ^{\sigma }\right) ^{\dagger }\left( \gamma ^{\rho }\right)
^{\dagger }.  \label{B2}
\end{equation}%
Hence%
\begin{equation}
\left( \gamma ^{\sigma }\right) ^{\dagger }\left( \gamma ^{\rho }\right)
^{\dagger }\gamma ^{\nu }\gamma ^{\lambda }=\gamma ^{0}\gamma ^{\sigma
}\gamma ^{\rho }\gamma ^{0}\gamma ^{\nu }\gamma ^{\lambda },  \label{B3}
\end{equation}%
and%
\begin{equation}
\gamma ^{\nu }\gamma ^{\lambda }\left( \gamma ^{\sigma }\right) ^{\dagger
}\left( \gamma ^{\rho }\right) ^{\dagger }=\gamma ^{\nu }\gamma ^{\lambda
}\gamma ^{0}\gamma ^{\sigma }\gamma ^{\rho }\gamma ^{0}.  \label{B4}
\end{equation}%
Since $\gamma ^{0}\gamma ^{\nu }\gamma ^{\lambda }\gamma ^{0}$ is the
Hermitian conjugate of $\gamma ^{\nu }\gamma ^{\lambda }$, the product $%
\gamma ^{0}\gamma ^{\sigma }\gamma ^{\rho }\gamma ^{0}$ commutes in the same
pattern with $\gamma ^{\nu }\gamma ^{\lambda }$, when surrounded by $\gamma
^{0}$. This is a general symmetry of Dirac matrices, because the $\gamma
^{\mu }$ matrices form a complete basis and their Hermitian conjugate simply
rotate is representation space. Hence the reordering is valid. In other words%
\begin{equation}
\left( \gamma ^{\sigma }\right) ^{\dagger }\left( \gamma ^{\rho }\right)
^{\dagger }\gamma ^{\nu }\gamma ^{\lambda }=\gamma ^{\nu }\gamma ^{\lambda
}\left( \gamma ^{\sigma }\right) ^{\dagger }\left( \gamma ^{\rho }\right)
^{\dagger }.  \label{B5}
\end{equation}

\end{document}